 \documentclass[final,1p,times]{elsarticle}
\usepackage{graphicx}
\usepackage{stfloats}

\usepackage{amssymb}
\usepackage{amsmath}
 
\newcommand{\rate}[2]{\ensuremath{k^{#1}_{#2}}}

\newcommand{\bmat}{\left(\begin{smallmatrix}}
\newcommand{\emat}{\end{smallmatrix}\right)}
\newcommand{\Rempty}[1]{\ensuremath{\textrm{#1}}}
\newcommand{\Rdna}[1]{\ensuremath{\bar{\textrm{#1}}}}
\newcommand{\Rligand}[1]{\ensuremath{\textrm{#1}^*}}
\newcommand{\Rbound}[1]{\ensuremath{\bar{\textrm{#1}}^*}}
\newcommand{\ligmat}[1]{\ensuremath{L_{\textrm{#1}}}}
\newcommand{\dnamat}[2]{\ensuremath{D^{\textrm{#1}}_{\textrm{#2}}}}
\newcommand{\Rep}[1]{\(R_{\textrm{#1}}\)}
\newcommand{\Diff}{\ensuremath{\tau_\textrm{diff}}}
\newcommand{\Lig}{\ensuremath{\tau_\textrm{lig}}}
\newcommand{\Holo}{\ensuremath{1/k^{\textrm{holo}}_{\textrm{off}}}}

\journal{Chemical Physics}

\begin{document}

\begin{frontmatter}

%% Title, authors and addresses

%% use the tnoteref command within \title for footnotes;
%% use the tnotetext command for theassociated footnote;
%% use the fnref command within \author or \address for footnotes;
%% use the fntext command for theassociated footnote;
%% use the corref command within \author for corresponding author footnotes;
%% use the cortext command for theassociated footnote;
%% use the ead command for the email address,
%% and the form \ead[url] for the home page:
%% \title{Title\tnoteref{label1}}
%% \tnotetext[label1]{}
%% \author{Name\corref{cor1}\fnref{label2}}
%% \ead{email address}
%% \ead[url]{home page}
%% \fntext[label2]{}
%% \cortext[cor1]{}
%% \address{Address\fnref{label3}}
%% \fntext[label3]{}

\title{A Master equation approach to modeling an artificial protein motor}
	
%% use optional labels to link authors explicitly to addresses:
%% \author[label1,label2]{}
%% \address[label1]{}
%% \address[label2]{}

\author[label1]{Nathan J. Kuwada\corref{label4}}
\author[label1,label2]{Gerhard A. Blab}
\author[label1,label3]{Heiner Linke}
\address[label1]{Department of Physics and Materials Science Institute, University of Oregon, 1274 University of Oregon, Eugene, OR, 97403-1274, USA}
\address[label2]{Department of Physics and IRMACS Centre, Simon Fraser University, Burnaby, British Columbia, V5A 1S6, Canada}
\address[label3]{The Nanometer Structure Consortium and Division of Solid State Physics, Lund University, Box 118, 22100 Lund, Sweden}
\cortext[label4]{Corresponding author, Tel 1-541-346-4583 Email address: nkuwada@uoregon.edu}
\begin{abstract}

Linear bio-molecular motors move unidirectionally along a track by coordinating several different processes, such as fuel (ATP) capture, hydrolysis, conformational changes, binding and unbinding from a track, and center-of-mass diffusion. A better understanding of the interdependencies between these processes, which take place over a wide range of different time scales, would help elucidate the general operational principles of molecular motors. Artificial molecular motors present a unique opportunity for such a study because  motor structure and function are \textit{a priori} known.  Here we describe use of a Master equation approach, integrated with input from Langevin and molecular dynamics modeling, to stochastically model a molecular motor across many time scales. We apply this approach to a specific concept for an artificial protein motor, the Tumbleweed.  

\end{abstract}

\begin{keyword}
%% keywords here, in the form: keyword \sep keyword
Artificial Molecular Motors \sep Master equation \sep Computational Simulation
%% PACS codes here, in the form: \PACS code \sep code

%% MSC codes here, in the form: \MSC code \sep code
%% or \MSC[2008] code \sep code (2000 is the default)

\end{keyword}

\end{frontmatter}

%\setpagewiselinenumbers
%\modulolinenumbers[5]
%\linenumbers

%% main text
\section{Introduction}
\label{sec:intro}

Naturally occurring protein-based molecular motors are a broad class of macromolecules that transduce chemical potential energy into directed transport. Examples include rotary motors responsible for ATP synthesis (ATP-synthase) and locomotion (flagella), as well as  linear motors, which move unidirectionally along a polymeric track and are responsible for, e.g., muscle contraction, cargo transport, and DNA/RNA transcription and replication~\cite{Howard01,Alberts08,Banting00}.  Enabled by revolutionary advances in single molecule detection techniques, such as fluorescence microscopy and optical tweezers, the activity and performance of individual molecular motors has been directly observed (for an overview of techniques and accomplishments see~\cite{Greenleaf07}).

Inspired by biomolecular motors, several efforts to construct artificial molecular motors have been pursued for the last decade or so, one driving force being the vision of biomimetic, nanoscale machinery~\cite{Feringa07}. In addition, artificial molecular motors offer an opportunity to develop better understanding of the general operational principles of molecular motors, because of the design freedom offered by synthetic motors, and because motor structure and function typically are \textit{a priori} known, allowing for detailed modeling. 

One approach to artificial molecular motors is the use of synthetic, small molecules, which has led to a large variety of designs of motors and motor parts~\cite{Kottas05, Balzini06, Kay07, Silvi08} including a surface-mounted, photochemically driven, rotary motor where a combination of light-driven \textit{cis-trans} isomerizations and thermal relaxations cause a molecule to make a complete 360\(^\circ\) rotation in a predefined direction~\cite{Feringa05}. This particular design has also been subject of detailed modeling~\cite{Klok09, Feringa09}, exemplifying this advantage of synthetic motors. 

In a competing approach~\cite{Yurke00}, and taking advantage of the highly designable, self-organized synthesis offered by oligonucleotides , Bath and Turberfield have constructed a linear DNA-based motor that is designed to step unidirectionally along a DNA-track~\cite{Bath09}.  The motor consists of two single-stranded (ss) DNA segments as `feet,' flexibly joined by a double-stranded (ds) linker, that attach to complementary sequences on an ss-track.  To achieve directional stepping, an ss-DNA fuel is introduced which selectively detaches the rear foot by competitive binding, allowing the rear foot to diffuse forward and complete a step. 

In a very recent third approach, a design concept for an artificial, protein-based motor was proposed~\cite{HFSP09}, with the aim to develop artificial motors that are based on the same material as biological motors. This design, the Tumbleweed, uses externally controlled, ligand-gated binding of repressor proteins to a DNA track to achieve unidirectional motion, and will be described in more detail below (Fig.~\ref{fig:steppingfig}). 

% INTRO TO MODELING
Artificial and biomolecular motors alike must coordinate many processes to achieve unidirectional motion. In the case of linear DNA- and protein motors, these processes generally include fuel (ligand) capture, molecular conformational changes, binding to and unbinding from a track, and center-of-mass diffusion. The understanding of the interplay between these processes, which can span a vast range of time scales, is central to the design process of artificial motors, and to developing a better understanding of molecular motors in general.

Several different modeling approaches have been used to model molecular motors, each addressing different aspects of this many-time-scales problem.  First, fundamental approaches to directed motion of small, Brownian molecules were developed based an spatially asymmetric ratchets~\cite{Magnasco93, Bier94} and Brownian Motor models~\cite{Astumian97, Hanggi02}, and helped understand the role of symmetry breaking, thermal noise and diffusion in molecular motors~\cite{Astumian07}. Such models are typically not designed to capture the molecule-specific \emph{structure-function} relationship as well as details about sub-step processes of biological motors. 

Molecular dynamics (MD) and Langevin dynamics models (LD), where molecular trajectories are calculated from an equation of motion, provide single-molecule information with potentially high accuracy, but require large amounts of computational time to simulate motor stepping, because all short-time processes must be explicitly calculated during the simulation run. One example of this approach are recent LD models of myosin V~\cite{Craig09,Vilfan09,Vilfan05}, where the modeling of motor stepping was enabled by coarse-graining of the molecule. Fully atomistic molecular dynamics (MD) simulations are currently computationally limited to simulation times on the order of nanoseconds, and thus too demanding for full simulations of motor stepping. 

Finally, stochastic models, such as the Master equation approach described in more detail below, sacrifice the single-molecule information of dynamic models by modeling transitions between ensembles of motor states, but have the great advantage that they allow the modeling of motor processes across many different time scales~\cite{Bustamante01}.

In developing a complex artificial motor, one would thus like to combine the advantages of several of the above approaches, specifically the ability to relate atomistic detail to resulting motor performance in terms of stepping rate and run length. Here we describe the integration of Master equation (ME) modeling with results from LD and MD simulations in the design process of the protein motor Tumbleweed~\cite{HFSP09}. In the following, we first describe the Tumbleweed concept in more detail, followed by a discussion of the time scales of a number processes that must be mutually tuned for stepping to result. Some of these time scales, such as the time for diffusional search for the next binding site, depend on motor design, are thus tunable to a degree, and can be estimated from LD modeling. Using a ME approach, enabled by our detailed knowledge of the possible motor states in this designed motor, we can explore the sensitivity of the motor design to chosen times and rates. The model output, such as quantitative relationships that must be fulfilled for successful motor performance, can then be used in the experimental motor design.

%%%
\section{The Tumbleweed Motor}
\label{sec:TW}

\begin{figure}[ht!]
\centering
\includegraphics {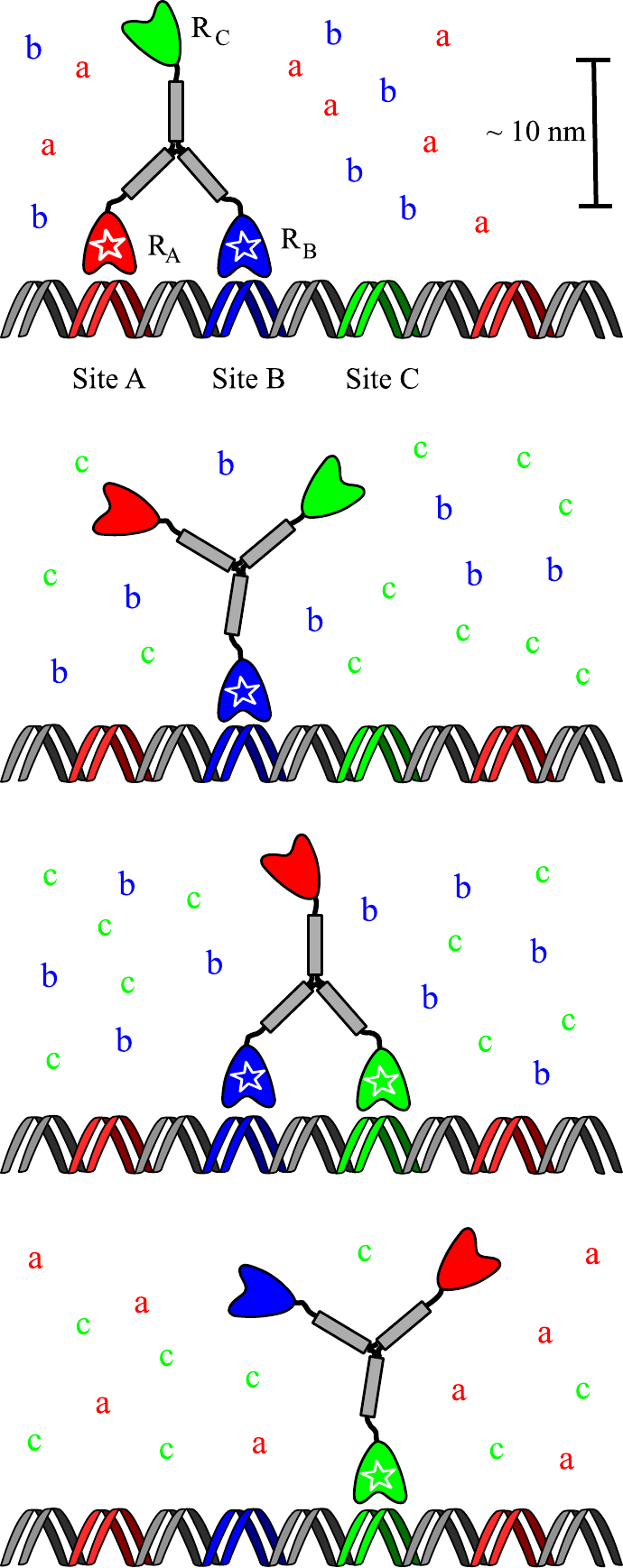}
\caption{Stepping process of the Tumbleweed. A coordinated, directed stepping of the TW motor is achieved by control of the ligand supply (letters a, b, and c), and DNA binding site arrangement. Each repressor protein foot (\Rep{A}, \Rep{B}, and \Rep{C}, shown in red, blue and green, respectively) is only able to bind strongly to its corresponding binding site while it has its specific ligand bound (indicated by a white star). Ligands are offered in a temporally repeating sequence ("plugs") [\textit{a,b}], [\textit{b,c}], [\textit{c,a}]. For the motor to successfully complete a step, the time scales for ligand capture, motor diffusion, and DNA binding must obey Eq. \ref{eq:inequal}.}
\label{fig:steppingfig}
\end{figure}

The Tumbleweed (TW) motor concept was developed as a first step towards artficial motors based on proteins~\cite{HFSP09}.  It consists of three  DNA-binding repressor proteins (\Rep{A}, \Rep{B}, \Rep{C}) attached to the corners of a Y-shaped, designed coiled-coil protein hub.  The repressor proteins (the motor `feet') have characteristically strong binding affinity to a unique ds-DNA base motif only in the presence of a specific ligand in solution (\textit{a}, \textit{b}, \textit{c}). Thus the binding activity of each foot can be controlled separately by the external ligand supply.  The track is then a designed length of ds-DNA with cyclic repeats of the three repressor protein binding motifs (binding sites).  Stepping of the TW molecule is by center-of-mass diffusion, and directed transport is achieved by coordinated binding and unbinding of the motor feet, externally coordinated by cycling ligands in solution.  The direction of transport is determined by both the binding site arrangement and the temporal order of the external ligand supply, and the speed of the motor is determined by how quickly ligands can be exchanged (see Fig.~\ref{fig:steppingfig}).

%%%%
\section{Modeling Tumbleweed: A Problem of Timescales}
\label{sec:timescales}

Consider TW in the state shown at the top of Fig.~\ref{fig:steppingfig}, with repressors \Rep{A} and \Rep{B} bound. To initiate a step, ligand \textit{a} in solution is replaced by ligand \textit{c}. For the step to take place, first the binding protein \Rep{A} must lose its ligand \textit{a}, transitioning from its holo- (with ligand) to its apo-state (without ligand), and protein \Rep{C} must gain its ligand \textit{c}. Apoprotein \Rep{A} must then unbind from the track, and the motor must diffuse until holoprotein \Rep{C} is near its next binding site and binds. Throughout this process, the external ligand supply must not be changed, and holoprotein \Rep{B} must stay attached to the track, or else the motor will fall off and be lost in the solution.

Qualitatively, these conditions are summarized by the inequality

\begin{equation}
\frac{1}{k_{\textrm{on/off}}} < \tau_{\textrm{diff}} < \tau_{\textrm{lig}} < \frac {1}{k^\textrm{holo}_\textrm{off}},
\label{eq:inequal}
\end{equation} 

in which \(k_{\textrm{on/off}}\) are the ligand-repressor (dis)association rates, \(\tau_{\textrm{diff}}\) is the diffusional stepping (searching) time of the motor, \(\tau_{\textrm{lig}}\) is the time period during which the ligand concentration is kept constant, and \(k^\textrm{holo}_\textrm{off}\) is the rate for a repressor holoprotein detachment from the DNA track.

A central modeling question for the design of TW molecule and experimental setup is: What are these different time scales and how will their absolute and relative values affect the performance of Tumbleweed?

The value of 1/\(k_{\textrm{on/off}}\) is not well known for our choice of repressor proteins, but is thought to be relatively fast compared to any other time scales in the system, on the order of picoseconds~\cite{Schadd93} . Modeling the system on this level requires a full atomistic molecular dynamics approach (MD), where the motion and interactions of atoms in the molecule as well as the surrounding ligands and buffer solution are explicitly calculated.  The second term in Eq.~\ref{eq:inequal} is the characteristic time for the TW molecule to rotationally diffuse to it's next binding site.  To get a rough idea of this term, the time scale for a sphere with the approximate diameter of the TW molecule (\(\sim\) 20~nm) to diffuse one binding length (\(\sim\)10~nm) is around 1~\(\mu\)s.  To get a detailed value for this term, one can also use fully atomistic MD modeling, which allows one to consider steric hindraces between molecular components and to understand effective flexibilities of the hub-repressor protein joints. For current computational power, the full atomistic MD of the TW is able to provide maximum simulation runs of approximately \(10^{-9}\)~s, which is near the time scale for diffusion of subsections of the molecule, such as one arm of the central hub, but is not long enough to completely explore \(\tau_{\textrm{diff}}\).  To fully explore how the stepping time depends on parameters such as joint flexibility or molecule size, we can coarse-grain the MD model to decrease computational time and increase the total simulation run-length time.  Inertial motion of the TW molecule becomes completely damped by the surrounding fluid for times greater than \(10^{-9}\)~s, and thus can be neglected from the equations of motion for longer times.  The atomistic picture of the molecule interacting with fluid molecules can then be replaced with a Langevin Equation, in which the fluid-molecule interactions are introduced as stochastic thermal noise and where molecular components are approximated by geometric objects whose sizes and viscous drag coefficients match the original motor components.  This over-damped Langevin Dynamics (LD) approach can then use as input results from a MD model as described above, such as steric constraints and joint flexibilities, to correctly approximate the molecular components.  For TW, the LD approach is applicable to time scales between \(\mu\)s and ms, which is right around the center-of-mass diffusion time of \(\sim~150~\mu\textrm{s}\) \cite{HFSP09}.

The last two terms of Eq.~\ref{eq:inequal}, as well as the motor run lengths of interest, are in the 0.1-100~s range.  Neither one of the dynamical models mentioned are able to efficiently explore this time scale.  And because the dynamic models are tuned to model a specific range of time scales, they are not well equipped to explore the sensitivity of TW to the \emph{interactions} of processes across different time scales.  To quantitatively determine how sensitive TW is to the parameters in Eq.~\ref{eq:inequal}, which spans across many orders of magnitude, we must sacrifice single-molecule information and instead use a stochastic modeling technique that is able to model processes across many time scales: The Master equation.

%%%

\section{The Master Equation Approach}
\label{sec:mastereq}  
  
The basis of the Master equation (ME) approach is the identification of motor states and their associated transition rates. In the case of the Tumbleweed, motor states are defined as distinct combinations of binding between the motor, its ligands, and the DNA track.  For example, the motor state \Rligand{A}\Rempty{B}\Rbound{C} represents the situation where foot \Rep{A} has its ligand bound but is not attached to the DNA, foot \Rep{B} has no ligand bound, and foot \Rep{C} has both its ligand bound and is attached to the DNA track (~${}^*$~and ~$\bar{}$~ thus represent ligand and DNA binding, respectively).  In total, there are 80 distinct states for the TW motor: $2^3=8$~ ligand  binding states times 10 DNA-binding states, shown enumerated in Table \ref{tab:eightystates} below. The number of possible DNA binding states is larger than the number of ligand binding states, as there are three theoretically possible configurations if all three feet are bound at the same time.

\begin{figure}[h]
\begin{center}
\includegraphics{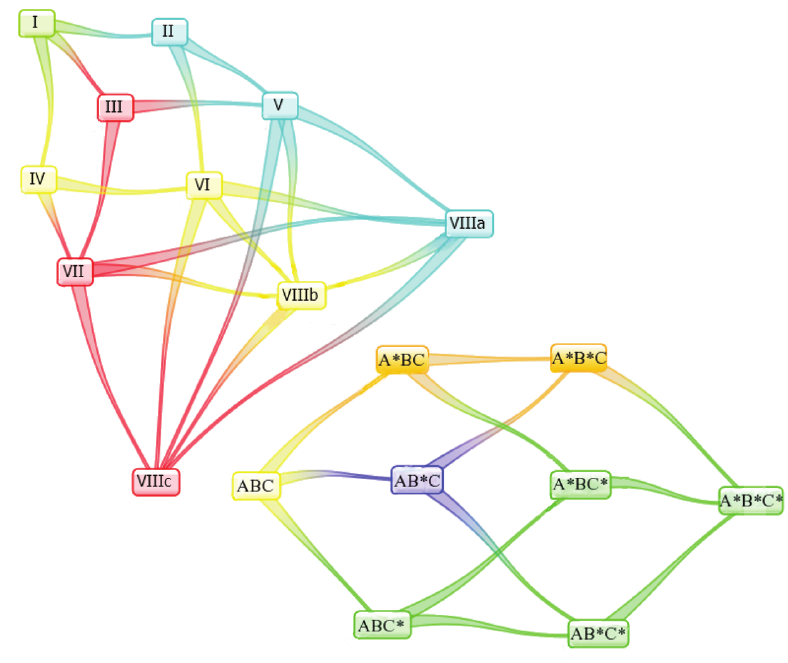}
\end{center}
\caption{Visual representation of the allowed transitions of the Tumbleweed motor. An allowed transition can either change the DNA binding state (indicated by roman numerals, top) or the ligand binding configuration (bottom, a star indicates the respective repressor has its ligand bound). A complete list of states is given in Table \ref{tab:eightystates}}
\label{fig:transitions}
\end{figure}

The ME is a differential equation that describes transitions between states. If all kinetics are approximated as first-order reactions, the ME is also of first order:

\begin{equation}
\frac{\partial}{\partial t} p(t)=M(t)p(t)
\label{eq:mastereq}
\end{equation}

where (for a system with N-states) \(p(t)\) is a N-dimensional vector with the numeric value of the n'th row representing the probability of finding the TW in the n'th motor state, and \(M(t)\) is an NxN matrix of transition rates between states.  A schematic representation of the allowed transitions for the TW motor is shown  in Fig.~\ref{fig:transitions}.  With the allowed transitions between the states defined, we can now form the \(M(t)\) transition matrix.  As shown in the state-naming convention of Table \ref{tab:eightystates}, it becomes convenient to define separate 8x8 sub-matrices \(L_{i}\) describing ligand exchange at constant DNA binding $i$, and \(D_{j}^i\), which represent a change between DNA binding states $i$ and $j$, respectively. The full \(M(t)\) is then: 

%% BIG M(t) MATRIX 

\begin{equation}
\bmat 
\ligmat{I} & \dnamat{I}{II} & \dnamat{I}{III} & \dnamat{I}{IV} & 0 & 0 & 0 & 0 & 0 & 0 \\
\dnamat{II}{I} & \ligmat{II} & 0 & 0 & \dnamat{II}{V} & \dnamat{II}{VI} & 0 & 0 & 0 & 0\\
\dnamat{III}{I} & 0 & \ligmat{III} & 0 & \dnamat{III}{V} & 0 & \dnamat{III}{VII}  & 0 & 0 & 0 \\
\dnamat{IV}{I} & 0 & 0 & \ligmat{IV} & 0 & \dnamat{IV}{VI} & \dnamat{IV}{VII} & 0 & 0 & 0 \\
0 & \dnamat{V}{II} & \dnamat{V}{III} & 0 & \ligmat{V} & 0 & 0 & \dnamat{V}{VIIIa} & \dnamat{V}{VIIIb} & \dnamat{V}{VIIIc} \\
0 & \dnamat{VI}{II}&  0 & \dnamat{VI}{IV} & 0 & \ligmat{VI} & 0 & \dnamat{VI}{VIIIa} & \dnamat{VI}{VIIIb} & \dnamat{VI}{VIIIc} \\
0 & 0 & \dnamat{VII}{III} & \dnamat{VII}{IV} & 0 & 0 & \ligmat{VII} & \dnamat{VII}{VIIIa} & \dnamat{VII}{VIIIb} & \dnamat{VII}{VIIIc} \\
0 & 0 & 0 & 0 & \dnamat{VIIIa}{V}  & \dnamat{VIIIa}{VI} & \dnamat{VIIIa}{VII} & \ligmat{VIIIa}& 0 & 0 \\
0 & 0 & 0 & 0 & \dnamat{VIIIb}{V}  & \dnamat{VIIIb}{VI} & \dnamat{VIIIb}{VII} & 0 & \ligmat{VIIIb} & 0\\
0 & 0 & 0 & 0 & \dnamat{VIIIc}{V}  & \dnamat{VIIIc}{VI} & \dnamat{VIIIc}{VII} & 0 & 0 & \ligmat{VIIIc} 
\emat
\label{eq:eightystate}
\end{equation}

with the two types of sub-matrices defined as:

 %% D_ij MATRIX
\begin{eqnarray}
\dnamat{i}{j}&=&
\bmat
\rate{ij}{1}&0&0&0&0&0&0&0 \\
0&\rate{ij}{2}&0&0&0&0&0&0 \\
0&0&\rate{ij}{3}&0&0&0&0&0 \\
0&0&0&\rate{ij}{4}&0&0&0&0 \\
0&0&0&0&\rate{ij}{5}&0&0&0 \\
0&0&0&0&0&\rate{ij}{6}&0&0 \\
0&0&0&0&0&0&\rate{ij}{7}&0 \\
0&0&0&0&0&0&0&\rate{ij}{8} \\
\emat
\label{eq:dnamat}
\end{eqnarray}

where \rate{ij}{\ell} describes the transition between two DNA binding configurations $i$ and $j$, while keeping the ligand binding configuration $\ell$ constant, and:
	
%% L_i MATRIX
\begin{equation}
L_i=
\bmat
-\Sigma & \rate{i}{-A|A} & \rate{i}{-B|B} & \rate{i}{-C|C} & 0 & 0 & 0 & 0 \\
\rate{i}{+A|0} & -\Sigma & 0 & 0 & \rate{i}{-B|AB} & \rate{i}{-C|AC} & 0 & 0 \\
\rate{i}{+B|0} & 0 & -\Sigma & 0 & \rate{i}{-A|AB} & 0 & \rate{i}{-C|BC} & 0 \\
\rate{i}{+C|0} & 0 & 0 & -\Sigma & 0 & \rate{i}{-A|AC} & \rate{i}{-B|BC} & 0 \\
0 & \rate{i}{+B|A} & \rate{i}{+A|B} & 0 & -\Sigma & 0 & 0 & \rate{i}{-C|ABC} \\
0 & \rate{i}{+C|A} & 0 & \rate{i}{+A|C} & 0 & -\Sigma & 0 & \rate{i}{-B|ABC} \\
0 & 0 & \rate{i}{+C|B} & \rate{i}{+B|C}& 0 & 0 & -\Sigma & \rate{i}{-A|ABC} \\
0 & 0 & 0 & 0 & \rate{i}{+C|AB} & \rate{i}{+B|AC} & \rate{i}{+A|BC} & -\Sigma
\emat
\label{eq:ligmat}
\end{equation}

\begin{table*}[b]
%\resizebox{1 \textwidth}
\center
{ \begin{tabular}{lc|cccccccccc}
&&I&II&III&IV&V&VI&VII&VIIIa&VIIIb&VIIIc \\
&&\Rempty{A}\Rempty{B}\Rempty{C}&\Rdna{A}\Rempty{B}\Rempty{C}&
\Rempty{A}\Rdna{B}\Rempty{C}&\Rempty{A}\Rempty{B}\Rdna{C}&
\Rdna{A}\Rdna{B}\Rempty{C}&\Rdna{A}\Rempty{B}\Rdna{C}&
\Rempty{A}\Rdna{B}\Rdna{C}&\Rdna{A}\Rdna{B}\Rdna{C}&
\Rdna{B}\Rdna{C}\Rdna{A}&\Rdna{C}\Rdna{A}\Rdna{B} \\ \hline
1&\Rempty{A}\Rempty{B}\Rempty{C}&1&9&17&25&33&41&49&57&65&73\\
2&\Rligand{A}\Rempty{B}\Rempty{C} \\
3&\Rempty{A}\Rligand{B}\Rempty{C} 
&$\downarrow$&$\downarrow$&$\downarrow$&$\downarrow$&$\downarrow$
&$\downarrow$&$\downarrow$&$\downarrow$&$\downarrow$&$\downarrow$ \\
4&\Rempty{A}\Rempty{B}\Rligand{C} \\
5&\Rligand{A}\Rligand{B}\Rempty{C} \\
6&\Rligand{A}\Rempty{B}\Rligand{C} 
&$\downarrow$&$\downarrow$&$\downarrow$&$\downarrow$&$\downarrow$
&$\downarrow$&$\downarrow$&$\downarrow$&$\downarrow$&$\downarrow$ \\
7&\Rempty{A}\Rligand{B}\Rligand{C} \\
8&\Rligand{A}\Rligand{B}\Rligand{C}&8&16&24&32&40&48&56&64&72&80
\end{tabular}}
\caption[Naming of the Tumbleweed States]{Naming convention for the 80 distinct states of the Tumbleweed motor. Roman numerals indicate states with identical DNA-binding (10 types), while arabic numbers on the left indicate identical ligand binding (8 types). Alternatively the states can also be numbered consequtivey from 1 to 80.}
\label{tab:eightystates}
\end{table*}

where \rate{i}{-B|AB} is the associated rate for the process \Rligand{A}\Rligand{B}C \(\Rightarrow\)  \Rligand{A}BC for the sustained DNA binding configuration $i$.  The term \(-\Sigma\) in the \(L_i\) matrix is the negative sum over the corresponding column of the full matrix, which is included to conserve the population.  Each of the 10 \(L_i\) matrices contains 24 independent rates, and each of the 36 \(D_j^i\) matrices is fully defined by 8 independent rates, which results in a maximal set of 528 independent parameters to completely describe the dynamics of the TW motor. Despite the large number of parameters in the TW Master equation, the actual transition matrix $M(t)$ is actually rather sparsely populated and is best treated as such to allow for efficient and fast calculation. A strength of the ME approach is the inclusion of the wide range of relevant time scales in a single calculation. This strength, however, results in a stiff equation, i.e.\ the equation includes terms that can lead to rapid variation and numerical instability with most standard solvers, unless the time steps are chosen extremely small. We have chosen a specialized solver for stiff ordinary differential equations (\texttt{ode5r}, GNU Octave~\cite{Eaton2002}) to allow us to solve the ME quickly and accurately.  

\section{Rates and States}
\label{sec:rates}

Many of the compound rates needed to populate the transition matrix Eq.~\ref{eq:ligmat} are based on the component rates of free repressor proteins, but we must allow for the possibility that they are modified by the sterical constraints and effective forces generated by the connecting hub and the binding to the DNA recogntion sites. 
Such modfications can at best be estimated by using LD or MD, but they are essentially not precisely known until a functioning prototype of the Tumbleweed motor has been constructed. For the moment we reduce the complexity of the problem by substituting those component rates known from literature for compound rates. In particular, we will consider the case where the ligand interaction $L_i$ does not depend on the current DNA binding state of the repressor, in essence reducing the set of parameters to the quantities introduced in Eq.~\ref{eq:inequal}. The most easily experimentally accessible parameters are the DNA binding and release times \rate{\mathrm{holo/apo}}{\mathrm{on}}~and \rate{\mathrm{holo/apo}}{\mathrm{off}}~in the presence (holo--) and absence (apo--) of ligand, respectively. While detailed information in the literature is sparse, \rate{\mathrm{holo}}{\mathrm{off}}~is generally found on the order of $5\cdot 10^{-3}~\mathrm{s}^{-1}$~\cite{Parsons1995,Stockley1998,Finucane2003,Wang2009}, and is expected to be experimentally tunable by salt concentration. The corresponding \rate{\mathrm{apo}}{\mathrm{off}}~is at least 2-3 orders of magnitude greater, with preliminary experimental results placing it even higher, on the order of $10^3~\mathrm{s}^{-1}$~\cite{Carey1988,Borden1991,Phillips1994}. The binding rate \rate{\mathrm{holo}}{\mathrm{on}}~depends on the relative concentration of repressor and DNA. In our simulations, we replace it by the diffusive search time \Diff~that we established by a Langevin simulation of the TW~\cite{HFSP09} and found to be about $\sim 150\mu$s. For the remainder of this paper, we use the following values unless specified other wise: \rate{\mathrm{holo}}{\mathrm{off}}~=~$10^{-2}~\mathrm{s}^{-1}$, \rate{\mathrm{apo}}{\mathrm{off}}~=~$10^{3}~\mathrm{s}^{-1}$, \Diff~=~$200~\mu s$, and \Lig~=~1~s.

%%%%

\section{Results}
\label{sec:results}

\begin{figure}[h]
\includegraphics{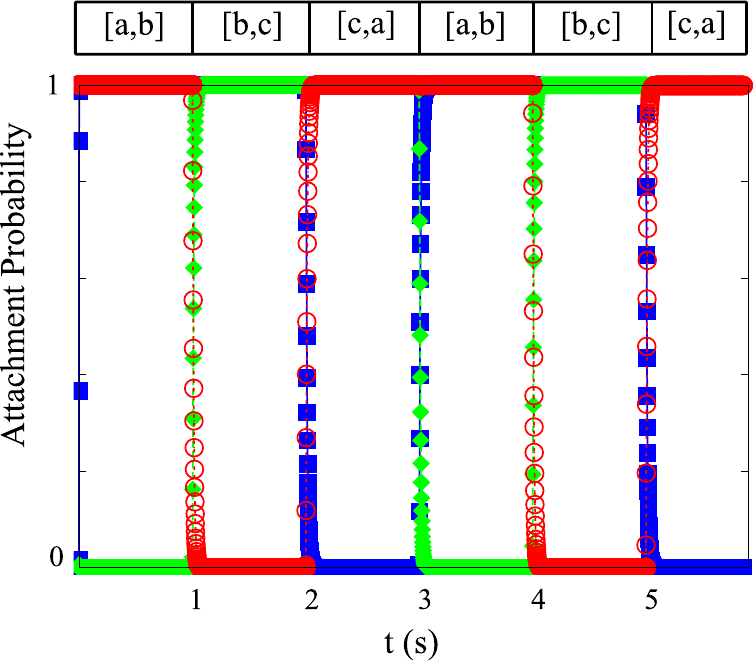}
\caption{Attachment of the TW feet (\Rep{A}: red circle, \Rep{B}: blue square, \Rep{C}: green diamond) to their specific DNA site over two complete ligand cycles. The ligand plugs are indicated over the graph. The values for \Diff~=~200~\(\mu\textrm{s}\) and the detachment rate \(1/k_{\textrm{off}}^{\textrm{holo}}~=~100~\textrm{s}\)~are chosen such that the motor is expected to complete successive steps. We can infer stepping motion by the binding behavior: during the entire [\textit{b,c}] ligand plug (for t = 1 s - 2 s), we see 100\% binding of \Rep{B} and \Rep{C}, while \Rep{A} is 0\%, but  \Rep{A} returns to 100\% binding probability during the following [\textit{c,a}]  plug (t = 2 s - 3 s) while \Rep{B} is 0\%, which is consistent with the ligand controlled stepping behavior. }
\label{fig:ABC}
\end{figure}

	A typical output from the Master equation model consists of the DNA binding probability for each foot as a function of time, as shown in Fig.~\ref{fig:ABC}.  The ME described here does not yield spatial information, such as a trajectory or the speed of the TW, but instead only determines in which binding state a molecule would most likely be.  Spatial information can be included in a ME simulation, but it requires one to define a unique set of states at each spatial coordinate, which will drastically increase the complexity of the problem and thus the computational time, depending on how long a track one wishes to include.  But we can imply stepping information from our ME by examining specific state transitions throughout the ligand exchange cycle. For example, if the binding probability for \Rep{A} is zero during the entire [\textit{b,c}] ligand plug and then returns to one during the [\textit{c,a}] plug, the motorÕs behavior is consistent with a state transition from state A\Rdna{B}\Rdna{C} to \Rdna{A}B\Rdna{C}, in agreement with the expected ligand controlled binding sequence \Rdna{A}\Rdna{B}C\(\rightarrow\)~A\Rdna{B}\Rdna{C}\(\rightarrow\)~\Rdna{A}B\Rdna{C}\(\rightarrow\)~\Rdna{A}\Rdna{B}C.    

We are now equipped to return to a question raised by Eq.~\ref{eq:inequal}: For Tumbleweed to take successive steps and not fall off the track, what is the required quantitative relationship between the diffusional step time (\Diff), the external ligand exchange time (\Lig) and the average time for the repressor proteins to detach from the track (\Holo)? 

\begin{figure}[h]
\includegraphics{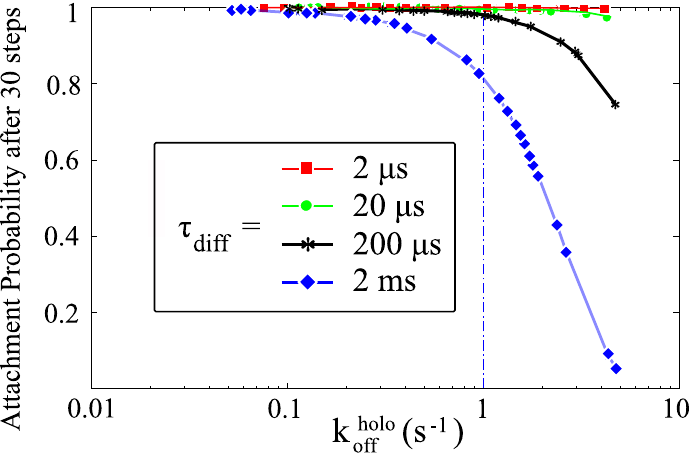}
\caption{Successive stepping, measured by the percentage of motor still attached to the DNA after 30 steps, for four values of diffusive search time (\Diff) as a function of detachment rate \(k^{\textrm{holo}}_{\textrm{off}}\). We find that the diffusive search time must be several orders of magnitude shorter than the average time for detachment for the motor to remain bound to DNA. For all curves the ligand plug length is constant (\Lig~=~1~s).}
\label{fig:taudiff}
\end{figure}

\Lig~is determined by the time scale on which the ligand concentration can be changed microfluidically, and can be expected to be on the order of 0.1-1~s. Fig.~\ref{fig:taudiff} shows the population of motors attached to the track after 30 steps as a function of \rate{\textrm{holo}}{\textrm{off}}~ for different values of \Diff~ with \Lig~ fixed at 1 s.  For \Diff~ on the 1- 100 $\mu$s time scale, the motor remains fairly successful even as \Holo~becomes less than \Lig, but for $\Diff~=~2~$ms, we see a 20\% reduction in motor attachment when \Holo~=~\Lig~=~1 s. It is initially surprising that the diffusive process can affect stepping success even though it is temporally separated from other relevant rates by 3 orders of magnitude.   A physical interpretation of this finding is as follows: the primary reason for the motor to detach from the track is when \Holo~ becomes less than \Lig~, because the motor is then simply not sticky enough. For short \Diff~however, the motor has a good chance to survive anyway, because it is attached to the track with two feet most of the time, providing a backup for one-foot detachment events.  However, the vulnerability for detachment is amplified for increasing \Diff, because the motor is attached with only one foot to the track for a greater fraction of the cycle.

\begin{figure}[h]
\includegraphics{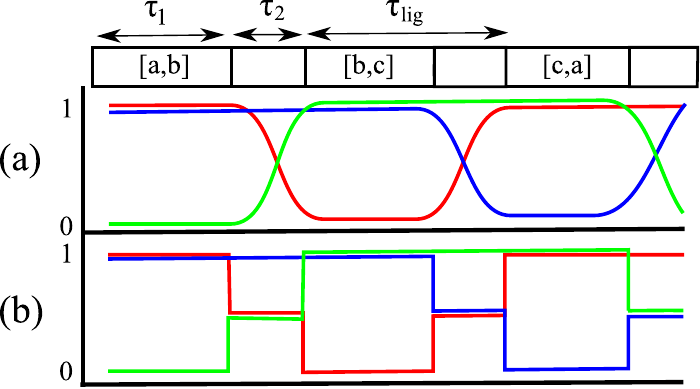}
\caption{Interdiffusion between clean, binary ligand plugs of duration (\Lig) results in the profile shown in (a), with time periods during which all three ligands (\textit{a, b}, and \textit{c}) are present in solution at the same time. For our numerical computations we use a simplified plug profile with discreet levels (b), characterized by a clean, binary plug of duration \(\tau_1\) ($<$ \Lig), and a mixed `pseudo-plug' of length \(\tau_2\). The ratio \(\tau_2\)/\Lig~defines the degree of the mixing.} 
\label{fig:ligmix}
\end{figure}

We can also use the ME approach to investigate experimental design considerations and their consequences on the performance of TW.   For example, there need to be at least three different buffer plugs containing ligands, [\textit{a},\textit{b}] [\textit{b},\textit{c}] [\textit{c},\textit{a}], microfluidically introduced in sequence into a microchamber containing the TW and immobilized DNA, to cycle the motor.  As there is a finite length the plugs must travel before they reach the molecules, there will be some interdiffusion between plugs, creating  pseudo-plugs containing all three ligands.  Although, in the current molecular design, the TW molecule is sterically hindered from binding all three binding proteins at the same time (state \Rdna{A}\Rdna{B}\Rdna{C} is not possible), the addition of triply mixed plugs reduces the temporal asymmetry of the system and could slow the motor down due to missteps.  How sensitive is the TW to plug-mixing, and does the experimental design need to be adjusted to compensate?  The expected experimental concentration profiles with interdiffusion between clean two-ligand plugs are shown qualitatively in Fig.~\ref{fig:ligmix}(a).  Although the ME is capable of modeling any ligand concentration profile, it is computationally expensive to model a continuous concentration function because the transition matrix \(M(t)\) has to be recalculated at each program time step.  Instead, we use the step-function approximated pseudo-plug profile shown in Fig.~\ref{fig:ligmix}(b), which is characterized by \(\tau_1\) and \(\tau_2\), the durations of the expected `clean' plug and the pseudo-plug, respectively, with \(\tau_1\)~+~\(\tau_2\)~=~\Lig. 

\begin{figure}[h]
\includegraphics{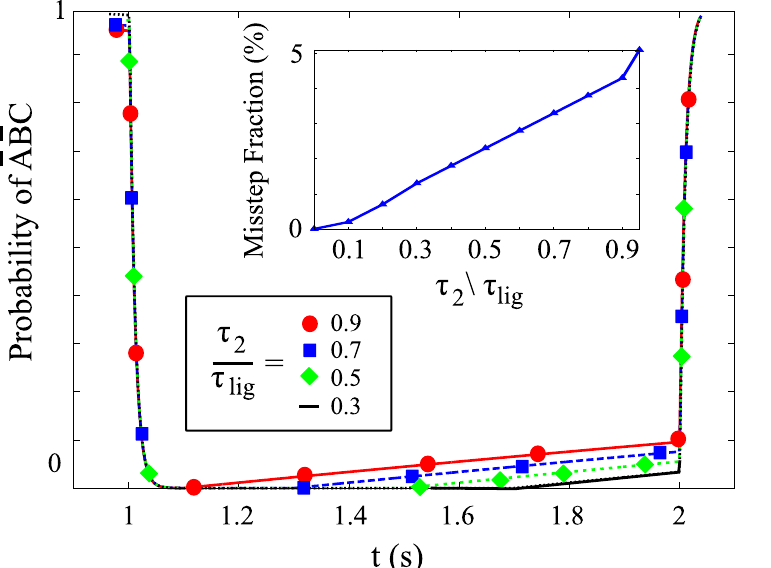}
\caption{ Attachment probability for state \Rdna{A}\Rdna{B}C during the [\textit{b,c}] ligand plug for four values of  \(\tau_2/\tau_{lig}\) with \Lig = 1 s.  The binding probability profile here implies that a fraction of the motors are either completing the diffusional step to the A\Rdna{B}\Rdna{C} state but then detaching \Rep{C} and returning to the previous \Rdna{A}\Rdna{B}C state, or simply not completing the step.  Inset: The percentage of motors that misstep when ligands nominally are exchanged  (time t = 2 s) as a function of \(\tau_2/\tau_{lig}\). }
\label{fig:mixcombo}
\end{figure}

Fig.~\ref{fig:mixcombo} shows the probability for state \Rdna{A}\Rdna{B}C during the [\textit{b,c}] ligand plug for multiple values of  \(\tau_2/\tau_{lig}\) with \Lig~= 1 s.  Instead of the 0\% binding probability for \Rep{A} throughout the entire [\textit{b,c}] plug characteristic for  \(\tau_2\) = 0 (Fig.~\ref{fig:ABC}), we now see a non-zero, increasing  probability for state \Rdna{A}\Rdna{B}C during the mixed pseudo-plug.  Because the \Rdna{A}\Rdna{B}C probability is 0\% during the unmixed portion of the [\textit{b,c}] plug \(\tau_1\), this profile implies two possible behaviors: (1) a fraction of motors completes the diffusional step forward to the A\Rdna{B}\Rdna{C} binding state but then detaches \Rep{C} and returns to the \Rdna{A}\Rdna{B}C state during \(\tau_2\); (2) motors release \Rep{A} but do not complete the step during \(\tau_1\) and return to state \Rdna{A}\Rdna{B}C.  In either scenario, these motors fall out of phase  with the ligand supply, resulting in the following binding sequence during the rest of the ligand cycle: \Rdna{A}\Rdna{B}C (original site)~\(\rightarrow\)~\Rdna{A}\Rdna{B}C~(misstep)~\(\rightarrow\)~\Rdna{A}B\Rdna{C}~(backstep to rearward site)~\(\rightarrow\)~\Rdna{A}\Rdna{B}C~(original site).   Therefore a single misstep causes the motor to temporally stall not just for a single ligand plug but for an entire ligand exchange cycle.  With \Lig~held constant, the longer \(\tau_2\), the less time the motors have to complete a step and the more opportunity they have to return to the previous binding state and stall.  The maximum percentage of these misstepping motors is determined by the fraction of motors in the \Rdna{A}\Rdna{B}C state at the ligand exchange time (t = 2 s in Fig.~\ref{fig:mixcombo}) and is shown in the inset of Fig.~\ref{fig:mixcombo}.

\begin{figure}[h]
\includegraphics{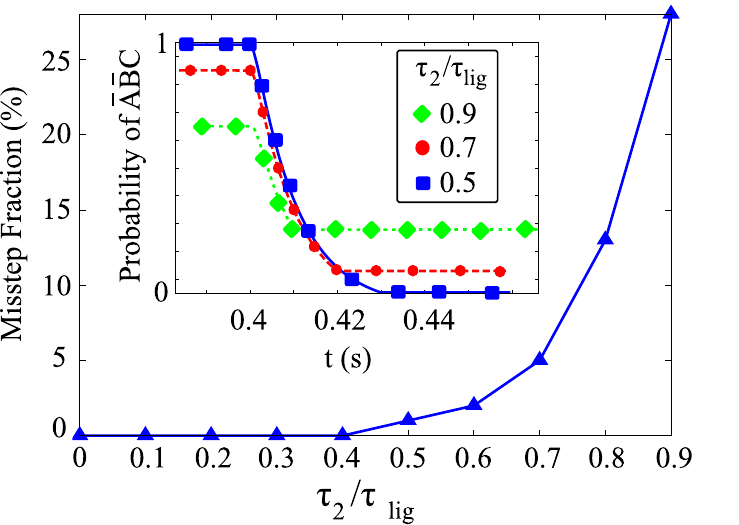}
\caption{Fraction of misstepping motors after ligand plug exchange [\textit{a},\textit{b}] to [\textit{b},\textit{c}] as a function of \(\tau_2/\tau_{lig}\) with \Lig~= 0.1 s .  Inset: Probability of state \Rdna{A}\Rdna{B}C for the [\textit{a},\textit{b}]  and [\textit{b},\textit{c}]  ligand plugs as a function of time for three values of  \(\tau_2/\tau_{lig}\) .  As the mixing time \(\tau_2\) is increased, the probability of the state \Rdna{A}\Rdna{B}C during the [\textit{b},\textit{c}] ligand plug no longer approaches zero and increases with higher \(\tau_2/\tau_{lig}\). Misstepping occurs when \(\tau_2\)/\Lig~$>$~0.5, where $\tau_2$ becomes comparable to 1/\rate{\mathrm{apo}}{\mathrm{off}}~(0.001 s) and a fraction of the motors never releases \Rep{A} before the mixed pseudo-plug arrives.  The high misstepping percentage also reduces the expected 100\% probability for state \Rdna{A}\Rdna{B}C during the [\textit{a,b}] ligand plug. }
\label{fig:ligAB}
\end{figure}

The maximum stepping rate of TW is determined by \Lig. What is the effect of ligand mixing if we chose to use a shorter \Lig~to increase the stepping rate? The inset of Fig. \ref{fig:ligAB} shows the probability for state \Rdna{A}\Rdna{B}C during the plugs [\textit{a,b}] and [\textit{b,c}] as a function of \(\tau_2/\tau_{lig}\) with \Lig~=~0.1 s, an order of magnitude less than Fig.~\ref{fig:mixcombo}.  Instead of the probability for state \Rdna{A}\Rdna{B}C going to 0\% during \(\tau_1\) as we saw in Figs.~\ref{fig:ABC} and~\ref{fig:mixcombo} for \(\tau_{lig}\) = 1~s, we now see that for \(\tau_2/\tau_{lig}~>~0.5\) a fraction of motors remain in state \Rdna{A}\Rdna{B}C throughout the entire pulse.  This effect occurs when \(\tau_1\) becomes similar to the characteristic fall-off time of the \Rdna{A}\Rdna{B}C state (\(\sim\)1/\rate{\mathrm{apo}}{\mathrm{off}}), so a fraction of motors have not yet released \Rep{A} before the mixed pseudo-plug arrives and thus remain in state \Rdna{A}\Rdna{B}C.  Because of the high misstepping percentage for \(\tau_2/\tau_{lig}~>~0.5\), we also begin to see a reduction in the probability for the expected state \Rdna{A}\Rdna{B}C during the [\textit{a,b}] ligand plug.  The misstepping fraction for \Lig~=~0.1~s is shown in Fig.~\ref{fig:ligAB}.  Therefore, although decreasing \Lig~has the possibility of increasing the speed of TW, the system also becomes more sensitive to high levels of mixing that could potentially \textit{decrease} the motor speed by increasing misstepping events.  Detailed ME modeling of these interdependencies will be highly useful for motor optimization.   

\section{Conclusions}
\label{sec:cons}

We have shown how a Master equation approach can be integrated with input from molecular dynamics and coarse-grained Brownian dynamics (Langevin) modeling to efficiently explore the dependence of the performance of a synthetic molecular motor on variables on a large range of time scales. 

For the specific case of the protein-motor concept Tumbleweed, returning to Eq. \ref{eq:inequal}, we have seen that the successive stepping success of TW is not remarkably sensitive to \Diff~ unless \Holo~ and \Lig~ are nearly equal.  In this case, for an experimentally realistic \Lig~=~1~s, the value of \Diff~ can decrease motor attachment by 20\% even if it is three orders of magnitude less than \Lig. Although the component holoprotein detachment rates (\(k_{\textrm{off}}^{\textrm{holo}}\)) are expected to be \(\sim\) $5\cdot 10^{-3}~\mathrm{s}^{-1}$ (suggesting that \Holo~$>>$ \Lig), the compound rates may vary due to steric constraints, and to be safe the design of the TW molecule should be tuned to decrease \Diff~ as much as possible.  Future MD modeling will be used to determine how design choices affect \Diff, e.g.\ through joint flexibilities and steric hindrances.

We have also determined how ligand plug mixing affects the stepping behavior of TW.  The binding state behavior suggests that plug mixing can lead to missteps and temporary stalling events if the motor returns to its previous binding state during the mixed plug time \(\tau_2\) and consequently falls out of phase with the periodic ligand supply.  Reducing \Lig~may, in principle, increase the speed of TW, but we have found the system also becomes more sensitive to ligand mixing, and for high levels of mixing temporary stalling events may actually decrease the speed, presenting an optimization challenge. For \Lig~=~1~s, a mixing time \(\tau_2/\tau_{lig}~=~0.5\) gives a misstepping probability of \(\sim 2.5\%\) per step.  Current experimental designs for the TW motor have the possibility of reducing \(\tau_2/\tau_{lig}\) to \(\sim 0.05\) where the probability of misstepping is negligible, thus the ligand mixing should not present a significant problem to the successful function of the Tumbleweed motor.  

The authors would like to thank the HFSP Motor Collaboration for inspiration and useful discussions, specifically Martin J. Zuckermann for technical advice.  This work is supported by the National Science Foundation under Grant Nos. DGE-0742540 and DGE-0549503 (NJK), and the Human Frontier Science Program (RGP0031/2007).

%% The Appendices part is started with the command \appendix;
%% appendix sections are then done as normal sections
%% \appendix

%% \section{}
%% \label{}

\end{document}